\documentclass[twocolumn,english,aps,prl,superscriptaddress,tightenlines]{revtex4-1}
\usepackage[T1]{fontenc}
\usepackage[latin9]{inputenc}
\usepackage{amsmath}
\usepackage{amssymb}
\usepackage{graphicx}
\usepackage{color}
\usepackage{hyperref}
\hypersetup{
    colorlinks=true, 
    linktoc=all,     
    linkcolor=blue,  
    citecolor=blue}
\usepackage{esint}

\makeatletter
\@ifundefined{textcolor}{}
{%
 \definecolor{BLACK}{gray}{0}
 \definecolor{WHITE}{gray}{1}
 \definecolor{RED}{rgb}{1,0,0}
 \definecolor{GREEN}{rgb}{0,1,0}
 \definecolor{BLUE}{rgb}{0,0,1}
 \definecolor{CYAN}{cmyk}{1,0,0,0}
 \definecolor{MAGENTA}{cmyk}{0,1,0,0}
 \definecolor{YELLOW}{cmyk}{0,0,1,0}
 }
\makeatother

\begin{document}

\title{Detector Readout of Analog Quantum Simulators}

\author{Lin Tian}

\affiliation{School of Natural Sciences, University of California, Merced, California 95343, USA}

\author{Iris Schwenk}

\affiliation{Institut f\"{u}r Theoretische Festk\"{o}rperphysik, Karlsruhe Institute of Technology (KIT), 76131 Karlsruhe, Germany}

\author{Michael Marthaler}

\affiliation{Institut f\"{u}r Theoretische Festk\"{o}rperphysik, Karlsruhe Institute of Technology (KIT), 76131 Karlsruhe, Germany}

\date{\today}

\begin{abstract}
An important step in quantum simulation is to measure the many-body correlations of the simulated model. For a practical quantum simulator composed of finite number of qubits and cavities, in contrast to ideal many-body systems in the thermodynamic limit, a measurement device can generate strong backaction on the simulator, which could prevent the accurate readout of the correlation functions. Here we calculate the readout of a detector coupled to an analog quantum simulator. We show that reliable characterization of the many-body correlations in the simulator can be achieved when the coupling operators obey the Wick's theorem. Our results are illustrated with two examples: a simulator for an harmonic oscillator and a simulator for the free electron gas. 
\end{abstract}
\maketitle

\emph{Introduction.}---Solving quantum many-body problems with classical methods often requires formidable computational resources. Quantum simulators, artificial systems constructed to mimic many-body models or dynamics, can solve such problems efficiently due to their built-in quantum parallelism~\cite{Feynman}. With the rapid experimental progresses, digital~\cite{LloydQS, digitalQS} and analog~\cite{Cirac+Zoller:12, HaukeReview} quantum simulators can be realized in various physical systems. Arrays of coupled qubits have been demonstrated with trapped ions~\cite{ion_exp1,ion_exp2, ion_review} and superconducting circuits~\cite{squbit_exp1,squbit_exp2, squbit_exp3, squbit_review}. Various equilibrium and out-of-equilibrium phenomena have been investigated with atoms and molecules in optical lattices~\cite{atom_review1,atom_review2}. Quantum simulation has been applied to problems in condensed-matter physics, quantum chemistry, nonequilibrium dynamics, and \emph{etc.}~\cite{AbramsPRL1997, LAWu2002, qchem1, eph1, eph2, eph3, fermihubbard, squbit_exp4, squbit_exp5}. 

The measurement of the temporal and spatial correlations in quantum simulators is a crucial component in analog quantum simulation~\cite{HaukeReview}. These correlators are often hard to calculate in computationally-demanding problems and can only be extracted from the readout of a detector. For ideal many-body systems in the thermodynamic limit, these correlators can be accurately determined from the detector's readout, where the detector perturbs the system weakly and the detector's backaction on the system can be safely neglected~\cite{Mahan}. For quantum simulators composed of finite number of qubits and cavities, the interaction between the simulator and the detector can distort the correlators of the simulator. It thus raises the question of whether the correlators of the simulator can be reliably extracted from the detector's readout. On the one hand, the detector relies on the existence of the simulator-detector coupling to retrieve the imprint of the correlators. On the other hand, the coupling generates a backaction that could affect the correlators of the simulated model. In \cite{LDuPRA2015}, the backaction of a cavity detector on a transverse field Ising model with $N$ sites was estimated with a perturbative approach and shows a $1/N$-dependence. Previously, quantum non-demolition measurement on many-body Hamiltonians was studied in atomic systems~\cite{BruunPRL2009, QND_exp}, which can only be applied to a limited group of simulators. 

Here we study the readout of a detector coupled to a quantum simulator with the Matsubara diagrammatic technique for finite temperature systems. The calculation is conducted on two examples: a simulator for a simple harmonic oscillator and a simulator for the free electron gas. We show that the unperturbed Matsubara correlators of the quantum simulator can be accurately extracted from the detector's readout when the coupling operators obey the Wick's theorem~\cite{WickT, Wick_1,Wick_2, Wick_3}. Whereas when this theorem is violated, the readout returns a correlator that is modified by the detector's backaction. In practice, Wick's theorem can be valid in models with gaussian fluctuations~\cite{Counting_1, Counting_2, Counting_3,Counting_4}. The imaginary-time Matsubara correlators can be transformed into real-time retarded correlators that are directly associated with the measurement via analytic continuation~\cite{Mahan, KlepfishNPB1998, WolfPRX2015}. Our result can thus be verified in experiments and can lead to the design of reliable measurement schemes for quantum simulators. We want to add that besides the detector's backaction, the reliability of analog quantum simulation can also be affected by noise and disorder~\cite{BrownPRL}. Several approaches to certifying quantum simulators under realistic conditions have been discussed, such as cross certification between simulators constructed with different systems and comparison with classical solutions in designated parameter regimes~\cite{HaukeReview, Leibfried, Johnson}. Note that in our treatment, the system-bath coupling is weak with its only effect being to thermalize the system to equilibrium. The influences of strong simulator-bath coupling and structured spectrum have been studied in \cite{Strong_Coupling_Quantum, Strong_Coupling_Schwenk, Strong_Coupling_Classical, wmzhang1, wmzhang2}.

\emph{Simulator and detector correlators.}---Consider a quantum simulator with Hamiltonian $H_{S}$ coupled to a detector with Hamiltonian $H_{R}$ via the interaction $H_{C}$. The total Hamiltonian of this system is $H_{T}=H_{S}+H_{R}+H_{C}$. The interaction can have the form $H_{C} = \sum_{i}\lambda_{i} \hat{O}_{S}(R_{i})\hat{\Gamma}_{R}$, where $\hat{O}_{S}(R_{i})$ ($\hat{\Gamma}_{R}$) is the coupling operator of the simulator (detector) at position $R_{i}$ and $\lambda_{i}$ is the site-dependent coupling constant ($\hbar=1$). The coupling of the simulator (detector) to bath modes is implicitly included in the Hamiltonians $H_{S}$ ($H_{R}$). The property of a many-body system can be revealed by its correlation functions. One Matsubara correlator for the simulator is defined as
\begin{equation}
{\cal C}_{S}(R_{i},R_{j}; \tau,\tau^{\prime})=-\left\langle \hat{T}_{\tau}\left[\hat{O}_{S}(R_{i}, \tau)\hat{O}_{S}(R_{j}, \tau^{\prime})\right]\right\rangle ,\label{eq:CSxtau}
\end{equation}
which describes the correlation between the coupling operators at imaginary times $\tau$ and $\tau^{\prime}$ with $\hat{O}_{S}(R_{i},\tau)=e^{H_{T}\tau}\hat{O}_{S}(R_{i})e^{-H_{T}\tau}$ and $\hat{T}_{\tau}$ being the time ordering operator. Written explicitly, ${\cal C}_{S}(R_{i},R_{j}; \tau,\tau^{\prime})=-{\rm Tr}[e^{-\beta H_{T}} \hat{T}_{\tau}[\hat{O}_{S}(R_{i},\tau)\hat{O}_{S}(R_{j},\tau^{\prime})]]/Z$ with $\beta=1/k_{B}T$ at temperature $T$ and $Z={\rm Tr}[e^{-\beta H_{T}}]$. For a system with translational symmetry, this correlator can be simplified to be ${\cal C}_{S}(R_{i}, \tau)$ by setting $R_{j}=0$ and $\tau^{\prime}=0$. The correlator ${\cal C}_{S}(R_{i}, \tau)$ can be transformed into the momentum and frequency domains as ${\cal C}_{S}(p, i\omega_{n})$ with quasimomentum $p$ and $\omega_{n}=2\pi n/\beta$ [$\omega_{n}=\pi(2n-1)/\beta$] for integer number $n$ when the operator $\hat{O}_{S}$ is bosonic [fermionic]~\cite{supple}. The correlator ${\cal C}_{S}(p,i\omega_{n})$ is the full correlator including the effect of the simulator-detector coupling. Without the coupling ($\lambda_{i}=0$), the correlator can be denoted as ${\cal C}_{S0}(p,i\omega_{n})$, which implicitly includes the effect of the simulator's bath modes. 
\begin{figure}
\includegraphics[width=8cm,clip]{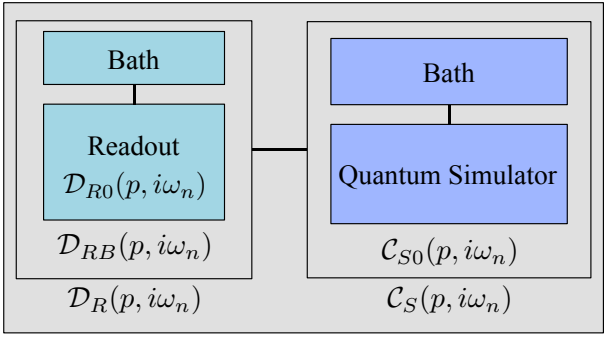}
\caption{The coupled simulator-detector system. The correlator of the bare simulator [detector] is denoted as ${\cal C}_{S0}(p,i\omega_{n})$ [${\cal D}_{R0}(p,i\omega_{n})$], the correlator of the detector with the bath modes is ${\cal D}_{RB}(p,i\omega_{n})$, and the full correlator of the simulator [detector] is ${\cal C}_{S}(p,i\omega_{n})$ [${\cal D}_{R}(p,i\omega_{n})$].}
\label{fig1}
\end{figure}

The correlator of the detector is defined as
\begin{equation}
{\cal D}_{R}(\tau,\tau^{\prime})=-\left\langle \hat{T}_{\tau}\left[\hat{\Gamma}_{R}(\tau)\hat{\Gamma}_{R}(\tau^{\prime})\right]\right\rangle \label{eq:DRxtau}
\end{equation}
with $\hat{\Gamma}_{R}(\tau)=e^{H_{T}\tau}\hat{\Gamma}_{R}e^{-H_{T}\tau}$. Under time invariance, this correlator can be simplified to be ${\cal D}_{R}(\tau)$, which transforms to ${\cal D}_{R}(i\omega_{n})$ in the frequency domain. Here ${\cal D}_{R}$ is the full correlator that includes the effects of the bath modes and the simulator. The coupling between the detector and the bath modes, e.g., the dissipation of a cavity detector, can often be engineered and characterized in experiments. The effect of the simulator-detector coupling on ${\cal D}_{R}$, when properly extracted from the readout, yields a measurement of the many-body correlations of the simulator. Denote the correlator of the bare detector as ${\cal D}_{R0}(p,i\omega_{n})$ and the correlator of the detector coupled to the bath modes (without the simulator) as ${\cal D}_{RB}(p,i\omega_{n})$. The relations of these three correlators are illustrated in Fig.~\ref{fig1}.

\emph{Diagrammatic expansion.}---The correlator of the detector is affected by the simulator-detector coupling. Using (\ref{eq:DRxtau}), we have ${\cal D}_{R}(\tau)=-\textrm{Tr}[e^{-\beta H_{T}}\hat{T}_{\tau}[\hat{\Gamma}_{R}(\tau)\hat{\Gamma}_{R}(0)]]/Z$ with $Z=\textrm{Tr}[e^{-\beta H_{T}}]$. The Matsubara diagrammatic technique can be applied to calculate this correlator~\cite{Mahan}. Let $U(\beta, 0) = e^{\beta H_{0}}e^{-\beta H_{T}}$ be the evolution operator from $\tau=0$ to $\tau=\beta$ in the interaction picture of Hamiltonian $H_{0}=H_{S}+H_{R}$. It can be shown that $U(\beta, 0)=\hat{T}_{\tau}[e^{-\int_{0}^{\beta}d\tau H_{CI}(\tau)}]$ with $H_{CI}(\tau)=e^{H_{0}\tau}H_{C}e^{-H_{0}\tau}$ being the coupling $H_{C}$ in the interaction picture. Then,
\begin{equation}
{\cal D}_{R}(\tau) = -\left\langle \hat{T}_{\tau}\left[U(\beta, 0)\hat{\Gamma}_{RI}(\tau)\hat{\Gamma}_{RI}(0)\right]\right\rangle _{0}/\left\langle U(\beta, 0)\right\rangle _{0} \label{eq:DR}
\end{equation}
with $\hat{\Gamma}_{RI}(\tau)=e^{H_{0}\tau}\hat{\Gamma}_{R}e^{-H_{0}\tau}$. Here the operator averages are taken on the thermal equilibrium of Hamiltonian $H_{0}$ with $\langle \cdots\rangle _{0}=\textrm{Tr}[e^{-\beta H_{0}}\cdots]/\text{Tr}[e^{-\beta H_{0}}]$. All the terms related to the coupling $H_{C}$ in (\ref{eq:DR}) are included in the evolution operator $U(\beta, 0)$. Treating $H_{CI}(\tau)$ as a perturbation, $U(\beta, 0)$ can be expanded into Taylor series of $H_{CI}(\tau)$. With the diagrammatic technique, the correlator ${\cal D}_{R}(\tau)$ can be written as a summation of diagrams in different orders of $H_{CI}(\tau)$~\cite{Mahan, supple}.

\emph{Cavity detector.}---Consider a cavity mode serving as the detector, which can be realized with optical or microwave cavities~\cite{cavityQED, circuitQED1, circuitQED2} and mechanical resonators~\cite{optomech1,optomech2}. The detector together with its bath modes has the Hamiltonian $H_{R}=\omega_{d}\hat{a}^{\dagger}\hat{a}+H_{B}$, where $\hat{a}$ ($\hat{a}^{\dagger}$) is the annihilation (creation) operator, $\omega_{d}$ is the frequency of the cavity mode, $H_{B}=\sum_{i} [c_{i}(\hat{b}_{i}+\hat{b}_{i}^{\dag})\hat{\Gamma}_{R}+H_{bi}]$ describes the cavity-bath coupling with $H_{bi}$ being the Hamiltonian of the $i$th bath mode, $\hat{\Gamma}_{R}=(\hat{a}+\hat{a}^{\dag})$ is the coupling operator of the cavity, and $\hat{b}_{i}$ ($\hat{b}_{i}^{\dagger}$) is the annihilation (creation) operator of the $i$th bath mode with coupling coefficient $c_{i}$. The spectrum of the bath modes is $J(\omega)=\pi\sum_{i}|c_{i}^{2}|\delta(\omega-\omega_{i})$~\cite{bathreview}. We assume the spectrum to be flat in the frequency range of interest with $J(\omega)=\kappa$ and $\kappa$ being the cavity bandwidth. The detector and the bath modes are in equilibrium at temperature $T$.

\begin{figure}
\includegraphics[clip,width=8cm]{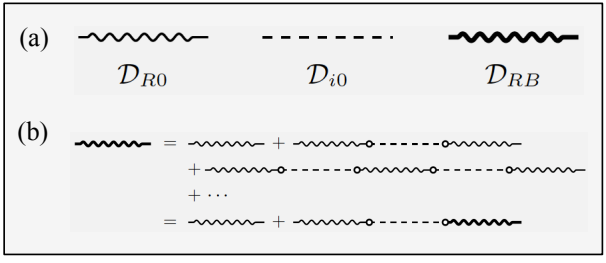}
\caption{(a) The diagrams for the correlators ${\cal D}_{R0}$ and ${\cal D}_{RB}$ of the cavity and ${\cal D}_{i0}$ of the $i$th bath mode. (b) The Dyson equation of ${\cal D}_{RB}$. The circles correspond to the coupling coefficients between the cavity and the bath modes.}
\label{fig2}
\end{figure}

We now derive the correlator of the detector coupled to the bath modes with a flat noise spectrum. This correlator will be utilized to calculate the detector's readout in the presence of a quantum simulator. The Matsubara correlator of the bare cavity (not coupled to the bath) can be derived from (\ref{eq:DRxtau}). In the frequency domain, the correlator has the form ${\cal D}_{R0}(i\omega_{n})=2\omega_{d}/[(i\omega_{n})^{2}-\omega_{d}^{2}]$ with $\omega_{n}=2\pi n/\beta$ and $n$ being an integer~\cite{supple}, where the temperature dependence enters the expression implicitly through the frequency $\omega_{n}$. Similarly, the correlator of an unperturbed bath mode is ${\cal D}_{i0}(i\omega_{n}) =2\omega_{i}/[(i\omega_{n})^{2}-\omega_{i}^{2}]$ with $\omega_{i}$ being the frequency of the $i$th bath mode. As illustrated in the diagrammatic equation in Fig.~\ref{fig2}, in the presence of the bath modes, the correlator of the cavity satisfies the Dyson equation
\begin{equation}
{\cal D}_{RB}={\cal D}_{R0}+{\cal D}_{R0}\left(\sum_{i}|c_{i}|^{2}{\cal D}_{i0}\right){\cal D}_{RB},\label{eq:DRB}
\end{equation}
where we omit the dependence on the variable $(i\omega_{n})$ in the correlators for brevity of discussion. The correlator of the cavity can then be obtained as ${\cal D}_{RB} = [1-{\cal D}_{R0}(\sum_{i}|c_{i}|^{2}{\cal D}_{i0})]^{-1}{\cal D}_{R0}$. Substituting ${\cal D}_{R0}$ and ${\cal D}_{i0}$ into ${\cal D}_{RB}$, we find that
\begin{equation}
{\cal D}_{RB}(i\omega_{n})=2\omega_{d}/\left[(i\omega_{n})^{2}-\widetilde{\omega}_{d}^{2}+i2\omega_{d}\kappa\right]\label{eq:DRBfinal}
\end{equation}
with $\widetilde{\omega}_{d}=(\omega_{d}^{2}+2\omega_{d}\delta\omega_{d})^{1/2}$ being the cavity frequency shifted by a small Cathy principle term $\delta\omega_{d}\ll\omega_{d}$~\cite{supple}. The correlator (\ref{eq:DRBfinal}) describes a cavity mode with frequency $\widetilde{\omega}_{d}\approx\omega_{d}$ and bandwidth $\sim\kappa$.

\begin{figure}
\includegraphics[clip,width=8cm]{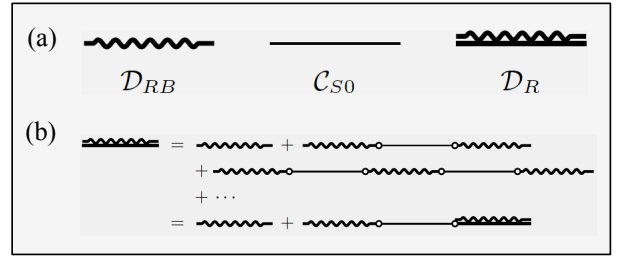}
\caption{(a) The diagrams for the correlators ${\cal D}_{RB}$ and ${\cal D}_{R}$ of the cavity and ${\cal C}_{S0}$ of the harmonic oscillator. (b) The Dyson equation of ${\cal D}_{R}$. The circles correspond to the coupling coefficient between the cavity and the oscillator. }
\label{fig3}
\end{figure}

\emph{Readout of oscillator mode.}---Here the simulator is for a simple harmonic oscillator with Hamiltonian $H_{S}=\omega_{s}\hat{a}_{s}^{\dag}\hat{a}_{s}$, where $\omega_{s}$ is the frequency and $\hat{a}_{s}$ ($\hat{a}_{s}^{\dag}$) is the annihilation (creation) operator of the oscillator. The coupling between the oscillator and the cavity detector has the form $H_{C}=\lambda\hat{O}_{S}\hat{\Gamma}_{R}$ with coupling operators $\hat{O}_{S}=(\hat{a}_{s}+\hat{a}_{s}^{\dag})$ and $\hat{\Gamma}_{R}=(\hat{a}+\hat{a}^{\dag})$, and coupling constant $\lambda$. Without the coupling ($\lambda=0$), the correlator of the operator $\hat{O}_{S}$ is ${\cal C}_{S0}(i\omega_{n})=2\omega_{s}/[(i\omega_{n})^{2}-\omega_{s}^{2}]$; the correlator of the operator $\hat{\Gamma}_{R}$ with the bath modes is given by (\ref{eq:DRBfinal}). With the coupling, the oscillator generates a force on the cavity that is proportional to the quadrature $\hat{O}_{S}$, which can be measured in the cavity's readout. The correlator of the cavity can be calculated with diagrammatic technique. As shown in Fig.~\ref{fig3}, the Dyson equation is 
\begin{equation}
{\cal D}_{R}={\cal D}_{RB}+|\lambda|^{2}{\cal D}_{RB}{\cal C}_{S0}{\cal D}_{R},\label{eq:DRoscillator}
\end{equation}
which yields ${\cal D}_{R}=(1-|\lambda|^{2}{\cal D}_{RB}{\cal C}_{S0})^{-1}{\cal D}_{RB}$. With ${\cal D}_{RB}$ in (\ref{eq:DRBfinal}), the full correlator of the cavity becomes ${\cal D}_{R}=[1-{\cal D}_{R0}(\sum_{i}|c_{i}|^{2}{\cal D}_{i0}+|\lambda|^{2}{\cal C}_{S0})]^{-1}{\cal D}_{R0}$. It can be viewed as a cavity coupled to a structured bath that contains the original bath modes and the oscillator. 

The correlator of the unperturbed oscillator can then be extracted from the measurement of the cavity correlators ${\cal D}_{RB}$ and ${\cal D}_{R}$ with
\begin{equation}
{\cal C}_{S0}=\left({\cal D}_{RB}^{-1}-{\cal D}_{R}^{-1}\right)/|\lambda|^{2}.\label{eq:DS0measure}
\end{equation}
This relation can be extended to a damped oscillator with finite bandwidth and can be used to characterize the parameters of the oscillator, such as its resonant frequency and bandwidth. Meanwhile, the simulator-detector coupling generates a backaction on the oscillator. The full correlator of the oscillator can be derived as ${\cal C}_{S}=(1-|\lambda|^{2}{\cal C}_{S0}{\cal D}_{RB})^{-1}{\cal C}_{S0}$. An interesting observation is that the readout of the cavity gives a faithful characterization of the unperturbed correlator ${\cal C}_{S0}$, instead of the full correlator ${\cal C}_{S}$, even though the coupling generates a finite backaction on the oscillator. This result indicates that finite detector backaction does not necessarily prevent the accurate characterization of a simulator.

\begin{figure}
\includegraphics[clip,width=8cm]{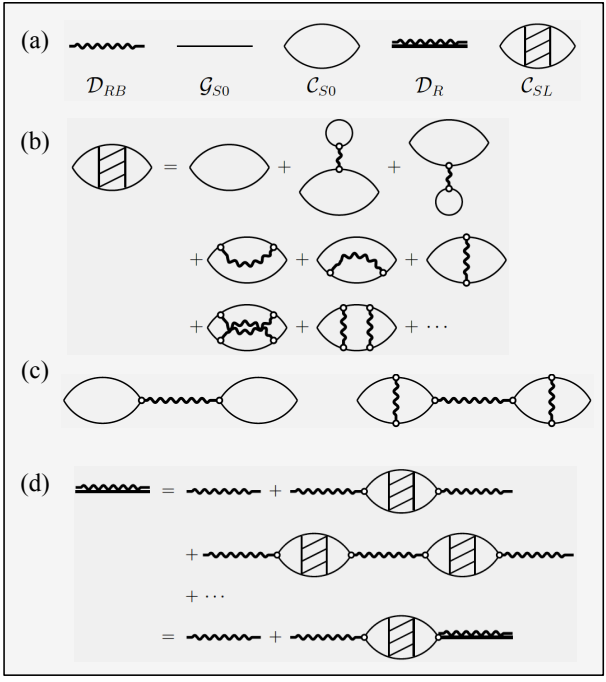}
\caption{(a) The diagrams for the correlators ${\cal D}_{RB}$ and ${\cal D}_{R}$ of the cavity, the propagator ${\cal G}_{0}$, the unperturbed correlator ${\cal C}_{S0}$, and the sum of the irreducible diagrams ${\cal C}_{SL}$ of the free electron gas. (b) ${\cal C}_{SL}$ with lower-order diagrams. (c) Examples of reducible diagrams. (d) The Dyson equation of ${\cal D}_{R}$. The circles correspond to the coupling coefficient between the cavity and the electrons.}
\label{fig4}
\end{figure}

\emph{Readout of free electron gas.}---The next simulator is for a free electron gas with Hamiltonian $H_{S}=\sum_{k}\epsilon_{k}\hat{c}_{k\sigma}^{\dagger}\hat{c}_{k\sigma}$, where $\epsilon_{k}$ is the energy and $\hat{c}_{k\sigma}$ ($\hat{c}_{k\sigma}^{\dag}$) is the annihilation (creation) operator of an electron with quasimomentum $k$ and spin $\sigma$. This model describes electrons in normal metals, where the Coulomb interaction between the electrons is largely screened by the interaction between the electrons and lattice ions. Assume that the interaction between the electron gas and the cavity detector has the form $H_{C}=\lambda\hat{O}_{S}(R_{0})\hat{\Gamma}_{R}$, where the coupling operator $\hat{O}_{S}(R_{0})=\sum_{k,q,\sigma}e^{iR_{0}q}\hat{c}_{k\sigma}^{\dag}\hat{c}_{(k+q)\sigma}$ is the density of the electron gas at position $R_{0}$~\cite{note1}. This interaction corresponds to the coupling between the electromagnetic field of the cavity and the electrons at $R_{0}$. Without the coupling ($\lambda=0$), the correlator of the operator $\hat{O}_{S}(R_{0})$ can be written as ${\cal C}_{S0}(i\omega_{n})=\sum_{k,q,\sigma}{\cal G}_{0\sigma}(k,i\omega_{n}){\cal G}_{0\sigma}(k+q,i\omega_{n})$, where ${\cal G}_{0\sigma}(k,i\omega_{n})=(i\omega_{n}-\epsilon_{k})^{-1}$ is the propagator of a free electron with quasimomentum $k$, imaginary frequency $i\omega_{n}$, and spin $\sigma$~\cite{supple}. The coupling generates a backaction on the simulator, which is manifested by various diagrams. Some of these diagrams are irreducible diagrams that cannot be disconnected by cutting off one propagator. We denote the sum of such diagrams as ${\cal C}_{SL}$ [Fig.~\ref{fig4}(a,b)]. For comparison, examples of reducible diagrams are given in Fig.~\ref{fig4}(c). 

The Dyson equation of the full correlator of the cavity can be written as
\begin{equation}
{\cal D}_{R}={\cal D}_{RB}+|\lambda|^{2}{\cal D}_{RB}{\cal C}_{SL}{\cal D}_{R},\label{eq:DRFEG}
\end{equation}
as shown by the diagrammatic equation in Fig.~\ref{fig4}(d). We then have ${\cal D}_{R}=(1-|\lambda|^{2}{\cal D}_{RB}{\cal C}_{SL})^{-1}{\cal D}_{RB}$. With the measured correlators ${\cal D}_{RB}$ and ${\cal D}_{R}$, we can obtain
\begin{equation}
{\cal C}_{SL}=\left({\cal D}_{RB}^{-1}-{\cal D}_{R}^{-1}\right)/|\lambda|^{2},\label{eq:CSLmeasure}
\end{equation}
i.e., the detector's readout returns a correlator modified by the detector's backaction.

\emph{Wick's theorem.}---For the simple harmonic oscillator, the readout of the cavity gives the unperturbed correlator ${\cal C}_{S0}$. Whereas the measurement of the free electron gas yields the correlator ${\cal C}_{SL}$ that is modified by the cavity's backaction. The distinct difference between these two cases can be explained by the validity of the Wick's theorem for the coupling operators in these models~\cite{WickT}. For demonstration, assume that the Wick's theorem is valid for the coupling operators $\hat{O}_{S}$ and $\hat{\Gamma}_{R}$ with
\begin{eqnarray}
&&\langle\hat{T}_{\tau}\hat{O}_{SI}(\tau_{1})\hat{O}_{SI}(\tau_{2})\cdots\hat{O}_{SI}(\tau_{2n-1})\hat{O}_{SI}(\tau_{2n})\rangle_{0} = \nonumber \\
 &  & \langle\hat{T}_{\tau}\hat{O}_{SI}(\tau_{1})\hat{O}_{SI}(\tau_{2})\rangle_{0}\langle\hat{T}_{\tau}\hat{O}_{SI}(\tau_{3})\cdots\hat{O}_{SI}(\tau_{2n})\rangle_{0}\nonumber \\
 &  & +\langle\hat{T}_{\tau}\hat{O}_{SI}(\tau_{1})\hat{O}_{SI}(\tau_{3})\rangle_{0}\langle\hat{T}_{\tau}\hat{O}_{SI}(\tau_{2})\cdots\hat{O}_{SI}(\tau_{2n})\rangle_{0}\nonumber \\
 &  & + \cdots, \label{eq:WT}
\end{eqnarray}
where $\hat{O}_{SI}(\tau)=e^{H_{0}\tau}\hat{O}_{S}e^{-H_{0}\tau}$ is the coupling operator in the interaction picture and we assume $\langle\hat{O}_{SI}(\tau)\rangle=0$. The correlators that contain $2(n-1)$ operators in (\ref{eq:WT}) will be further factorized using Wick's theorem. In the end, each term in the full correlator becomes a product of $n$ two-point correlators in the form of ${\cal C}_{S0}(\tau_{i}-\tau_{j})=-\langle\hat{T}_{\tau}\hat{O}_{SI}(\tau_{i})\hat{O}_{SI}(\tau_{j})\rangle_{0}$. Similar factorization can be performed on the correlators of the operator $\hat{\Gamma}_{R}$. Using (\ref{eq:WT}), the Dyson equation of the detector correlator can be derived as ${\cal D}_{R}={\cal D}_{R0}+|\lambda|^{2}{\cal D}_{R0}{\cal C}_{S0}{\cal D}_{R}$. The detector's readout can hence be used to characterize the unperturbed correlator of the simulator ${\cal C}_{S0}$.

The Wick's theorem is exactly valid when the coupling operator is a linear combination of the creation and annihilation operators of the quasiparticles in the unperturbed Hamiltonian $H_{0}$. For the harmonic oscillator, the coupling operator $\hat{O}_{s}=(\hat{a}_{s}+\hat{a}_{s}^{\dag})$ obeys the Wick's theorem. The density operator $\hat{O}_{S}(R_{0})$ of the free-electron gas, on the contrary, does not obey this theorem, and its correlator includes complicated diagrams resulting from the simulator-detector coupling. In realistic systems, the Wick's theorem can be valid for the coupling operators when the fluctuations in the system are approximately gaussian~\cite{Counting_1, Counting_2, Counting_3,Counting_4}. Under this approximation, the Dyson equation of the detector correlator depends on the unperturbed correlator of the simulator ${\cal C}_{S0}$. This enables the reliable readout of the correlator ${\cal C}_{S0}$ by the measurement device, which is the goal of quantum simulation. Furthermore, when ${\cal C}_{S0}$ is known, the detector's readout can be used to verify the validity of the Wick's theorem by comparing the measured correlator with ${\cal C}_{S0}$.

\emph{Discussions.}---We have studied the readout of a cavity detector coupled to a quantum simulator at given spatial location for simplicity of discussion. Our result can be applied to other types of detectors and other forms of coupling operators. For example, the measurement of spatial correlations is of wide interests in condensed-matter systems. By extending the coupling operator to include spatial dependence, i.e., $\hat{\Gamma}_{R}(R_{i})$, the impact of the detector's backaction on the readout of the spatial correlators can be studied. We can also study schemes that involve multiple detectors, each of which is coupled to the simulator via a distinct operator. 

In addition, our calculation is based on the imaginary-time Matsubara correlators. While the correlators measured in experiments are the retarded correlators defined in real time~\cite{Mahan, supple}. The retarded correlators can be obtained from the Matsubara correlators via analytic continuation. Various techniques have been developed to convert the Matsubara correlators to the retarded correlators numerically~\cite{KlepfishNPB1998, WolfPRX2015}.

\emph{Conclusions.}---To summarize, we studied the question of whether the readout of a detector can faithfully reflect the unperturbed many-body correlations in a quantum simulator using the Matsubara diagrammatic technique. Our result shows that when the coupling operators obey the Wick's theorem, the readout of the detector carries the imprint of the unperturbed correlator of the simulator and can be used to accurately characterize the correlations in the simulator. In contrast, when the Wick's theorem is violated, the measured simulator correlator is modified by the detector's backaction. Our work can shed light on the design of reliable measurement of quantum simulators and can lead to future endeavors in the proper characterization of quantum simulators.

\begin{acknowledgements}
L.T. is supported by the National Science Foundation under Award No. NSF-DMR-0956064. I.S. acknowledges financial support by Friedrich-Ebert-Stiftung.
\end{acknowledgements}

\newpage

\setcounter{equation}{0}
\setcounter{figure}{0}
\setcounter{table}{0}
\makeatletter
\renewcommand{\theequation}{S\arabic{equation}}
\renewcommand{\thefigure}{S\arabic{figure}}
\renewcommand{\bibnumfmt}[1]{[S#1]}
\renewcommand{\citenumfont}[1]{S#1}

\onecolumngrid

\begin{center}
\textbf{\large Supplemental Materials for ``Detector Readout of Analog Quantum Simulators''}
\end{center}

\section{I. Correlators in time and frequency domains}
The time-domain Matsubara correlator of the coupling operator $\hat{O}_{S}(R_{i})$ is defined as
\begin{equation}
{\cal C}_{S}(R_{i},R_{j}; \tau,\tau^{\prime})=-\left\langle \hat{T}_{\tau}\left[\hat{O}_{S}(R_{i}, \tau)\hat{O}_{S}(R_{j}, \tau^{\prime})\right]\right\rangle,\label{eq:CSxtau}
\end{equation}
where $\hat{O}_{S}(R_{i},\tau)=e^{H_{T}\tau}\hat{O}_{S}(R_{i})e^{-H_{T}\tau}$ is the operator at imaginary time $\tau$ with $-\beta< \tau\le\beta$, $\tau^{\prime}$ is an imaginary time, and $\hat{T}_{\tau}$ is the time ordering operator. We let $\hbar=1$ for simplicity of discussion. With translational symmetry, this correlator can be simplified as ${\cal C}_{S}(R_{i}, \tau)$ by setting $R_{j}=0$ and $\tau^{\prime}=0$. The momentum-domain counterpart of the correlator ${\cal C}_{S}(R_{i}, \tau)$ can be obtained by Fourier transformation and can be denoted as ${\cal C}_{S}(p, \tau)$. Note that when the operator $\hat{O}_{S}(R_{i})$ is bosonic, ${\cal C}_{S}(p, \tau)= {\cal C}_{S}(p, \tau+\beta)$ for $\tau<0$; when the operator $\hat{O}_{S}(R_{i})$ is fermionic, ${\cal C}_{S}(p, \tau)= -{\cal C}_{S}(p, \tau+\beta)$ for $\tau<0$~\cite{Mahan}. These relations are also valid for the spatial-domain correlator ${\cal C}_{S}(R_{i}, \tau)$. 

The frequency-domain counterpart of the time-domain correlator ${\cal C}_{S}(p, \tau)$ has the form: 
\begin{equation} 
{\cal C}_{S}(p, i\omega_{n}) = \int_{0}^{\beta} d\tau e^{i\omega_{n}\tau}{\cal C}_{S}(p, \tau), \label{eq:CSpomega}
\end{equation}
where $p$ is the quasimomentum and $\omega_{n}=2\pi n/\beta$ [$\omega_{n}=\pi(2n-1)/\beta$] for bosonic (fermionic) operator $\hat{O}_{S}(R_{i})$ with $n$ being an integer. The inverse transformation of (\ref{eq:CSpomega}) can be written as 
\begin{equation} 
{\cal C}_{S}(p, \tau) = \frac{1}{\beta}\sum_{n=-\infty}^{\infty}e^{-i\omega_{n}\tau} {\cal C}_{S}(p, i\omega_{n}).
\end{equation}

The correlator measured in experiments is the retarded correlator defined as
\begin{equation}
C_{S}^{ret}(R_{i},R_{j}; t, t^{\prime}) = -i\theta(t-t^{\prime})\left\langle\left[\hat{O}_{S}(R_{i},t)\hat{O}_{S}(R_{j},t^{\prime}) - \hat{O}_{S}(R_{j}, t^{\prime})\hat{O}_{S}(R_{i}, t)\right]\right\rangle,\label{eq:CSretxt}
\end{equation}
where $\theta$ is the Heaviside function, $\hat{O}_{S}(R_{i},t)=e^{iH_{T}t}\hat{O}_{S}(R_{i})e^{-iH_{T}t}$, and $t$, $t^{\prime}$ are real times. Here we have assumed that the coupling operator $\hat{O}_{S}(R_{i},t)$ is bosonic. For ferminonic operators, the retarded correlator is defined as $C_{S}^{ret}(R_{i},R_{j}; t, t^{\prime}) = -i\theta(t-t^{\prime})\langle[\hat{O}_{S}(R_{i},t)\hat{O}_{S}(R_{j},t^{\prime})+\hat{O}_{S}(R_{j}, t^{\prime})\hat{O}_{S}(R_{i}, t)]\rangle$. The retarded correlator in the momentum and frequency domains can be denoted as $C_{S}^{ret}(p,\omega)$. 

The retarded correlator can be obtained from the Matsubara correlator via analytic continuation: 
\begin{equation}
C_{S}^{ret}(p,\omega)={\cal C}_{S}(p,i\omega_{n}\rightarrow\omega+i\delta)\label{eq:eq:CSretanalytic}
\end{equation}
with $\delta$ being an infinitesimally small positive number~\cite{Mahan}. Because this relation is exact, the Matsubara correlators obtained in the main text can be directly converted to corresponding retarded correlators. Our result in the main text can hence be directly verified in experiments. Note that the conversion of a numerical Matsubara correlator to a retarded correlator via analytic continuation is nontrivial. Various techniques for such conversion have been developed~\cite{KlepfishNPB1998, WolfPRX2015}.

Note that the imaginary-time and real-time correlators of the coupling operator of the detector and their frequency counterparts can be defined similarly and will be discussed in the next section.

\section{II. Diagrammatic expansion of correlators}
The correlator of the detector is defined as
\begin{equation}
{\cal D}_{R}(\tau,\tau^{\prime})=-\left\langle \hat{T}_{\tau}\left[\hat{\Gamma}_{R}(\tau)\hat{\Gamma}_{R}(\tau^{\prime})\right]\right\rangle \label{eq:DRxtau}
\end{equation}
with $\hat{\Gamma}_{R}(\tau)=e^{H_{T}\tau}\hat{\Gamma}_{R}e^{-H_{T}\tau}$, where $H_{T}=H_{S}+H_{R}+H_{C}$ is the total Hamiltonian of this system, $H_{S}$ ($H_{R}$) is the Hamiltonian for the simulator (detector), and $H_{C}$ is the simulator-detector coupling. For a system with translational symmetry, the correlator can be simplified as ${\cal D}_{R}(\tau)$. Written explicitly, we have 
\begin{equation}
{\cal D}_{R}(\tau)=-\textrm{Tr}\left[e^{-\beta H_{T}}\hat{T}_{\tau}\left[\hat{\Gamma}_{R}(\tau)\hat{\Gamma}_{R}(0)\right]\right]/Z \label{eq:DRxtauFull}
\end{equation}
with $Z=\textrm{Tr}[e^{-\beta H_{T}}]$. Define the evolution operator $U(\beta, 0) = e^{\beta H_{0}}e^{-\beta H_{T}}$ in the interaction picture of the unperturbed Hamiltonian $H_{0}=H_{S}+H_{R}$. It can be shown that for $\tau>0$
\begin{eqnarray}
{\cal D}_{R}(\tau) &=& -\textrm{Tr}\left[e^{-\beta H_{0}} U(\beta, 0) U^{-1}(\tau, 0) \hat{\Gamma}_{RI}(\tau) U(\tau, 0) \hat{\Gamma}_{RI}(0)\right]/\textrm{Tr}\left[e^{-\beta H_{0}} U(\beta, 0)\right] \nonumber \\
&=& -\textrm{Tr}\left[e^{-\beta H_{0}} U(\beta, \tau)  \hat{\Gamma}_{RI}(\tau) U(\tau, 0) \hat{\Gamma}_{RI}(0)\right]/\textrm{Tr}\left[e^{-\beta H_{0}} U(\beta, 0)\right] \nonumber \\
&=& -\left\langle \hat{T}_{\tau}\left[U(\beta, 0)\hat{\Gamma}_{RI}(\tau)\hat{\Gamma}_{RI}(0)\right]\right\rangle _{0}/\left\langle U(\beta, 0)\right\rangle _{0},\label{eq:DR}
\end{eqnarray}
where $\hat{\Gamma}_{RI}(\tau)=e^{H_{0}\tau}\hat{\Gamma}_{R}e^{-H_{0}\tau}$ is the coupling operator $\hat{\Gamma}_{R}$ in the interaction picture. The operator averages in the third row of the above equation are taken over the thermal equilibrium of the Hamiltonian $H_{0}$ with $\langle \cdots\rangle _{0}=\textrm{Tr}[e^{-\beta H_{0}}\cdots]/\text{Tr}[e^{-\beta H_{0}}]$. In deriving (\ref{eq:DR}), we have used the relations $e^{-\beta H_{T}}=e^{-\beta H_{0}} U(\beta, 0)$, $e^{\beta H_{T}}=U^{-1}(\beta, 0)e^{\beta H_{0}} $, and $U(\beta, 0) U^{-1}(\tau, 0) = U(\beta, \tau)$.  

Let $H_{CI}(\tau)=e^{H_{0}\tau}H_{C}e^{-H_{0}\tau}$ be the simulator-detector coupling in the interaction picture. It can be shown that  
\begin{equation}
dU(\tau, 0)/d\tau = -  H_{CI}(\tau)U(\tau, 0), \label{eq:dU}
\end{equation}
which leads to 
\begin{equation}
U(\tau, 0)=\hat{T}_{\tau}[e^{-\int_{0}^{\tau}d\tau^{\prime} H_{CI}(\tau^{\prime})}]. \label{eq:Ubeta}
\end{equation}
Treating $H_{CI}$ as a perturbation, we can expand the operator $U(\beta, 0)$ as 
\begin{eqnarray}
U(\beta, 0) = 1-\int_{0}^{\beta}d\tau H_{CI}(\tau) + \frac{1}{2}\int_{0}^{\beta}d\tau \int_{0}^{\beta}d\tau^{\prime}\hat{T}_{\tau}\left[H_{CI}(\tau)H_{CI}(\tau^{\prime})\right] + \cdots.\label{eq:Ubeta2}
\end{eqnarray}
Substituting (\ref{eq:Ubeta2}) into (\ref{eq:DR}), we obtain the Taylor expansion of the detector correlator ${\cal D}_{R}$ in terms of the simulator-detector coupling~\cite{Mahan}. Similar derivation can be applied to the correlator of the simulator.

\section{III. Cavity detector}
For a bare cavity mode (without the bath modes and the simulator), the time-domain correlator (\ref{eq:DRxtauFull}) of the coupling operator $\hat{\Gamma}_{R}=(\hat{a}+\hat{a}^{\dag})$ can be derived as
\begin{equation}
{\cal D}_{R0}(\tau) =  -\theta(\tau)\left[e^{-\omega_{d}\tau}(n_{\textrm{th}}+1)+e^{\omega_{d}\tau}n_{\textrm{th}}\right] -\theta(-\tau)\left[e^{-\omega_{d}\tau}n_{\textrm{th}}+e^{\omega_{d}\tau}\left(n_{\textrm{th}}+1\right)\right],\label{eq:DRC}
\end{equation}
where $n_{\textrm{th}}=(e^{\beta\omega_{d}}-1)^{-1}$ is the thermal occupation number of the cavity mode at $\beta=1/k_{B}T$ (temperature $T$). Using the transformation ${\cal D}_{R0}(i\omega_{n}) = \int_{0}^{\beta} d\tau e^{i\omega_{n}\tau}{\cal D}_{R0}(\tau)$, we find the frequency-domain correlator:
\begin{equation}
{\cal D}_{R0}(i\omega_{n})=2\omega_{d}/\left[(i\omega_{n})^{2}-\omega_{d}^{2}\right],\label{eq:DR0}
\end{equation}
with $\omega_{n}=2\pi n/\beta$ and $n$ being an integer. As discussed in the main paper, the interaction between the cavity and its bath modes modifies the correlator of the cavity. We denote the correlator of the cavity mode affected by the bath modes as ${\cal D}_{RB}$. The diagrammatic equation generated by (\ref{eq:DR}) shows that 
\begin{eqnarray}
{\cal D}_{RB}&=&{\cal D}_{R0}+{\cal D}_{R0}\left(\sum_{i}|c_{i}|^{2}{\cal D}_{i0}\right){\cal D}_{R0} + {\cal D}_{R0}\left(\sum_{i}|c_{i}|^{2}{\cal D}_{i0}\right){\cal D}_{R0}\left(\sum_{i}|c_{i}|^{2}{\cal D}_{i0}\right){\cal D}_{R0}+\cdots \nonumber \\
&=&{\cal D}_{R0}+{\cal D}_{R0}\left(\sum_{i}|c_{i}|^{2}{\cal D}_{i0}\right){\cal D}_{RB}.\label{eq:DRB}
\end{eqnarray}
All correlators in the above equation have the variable $i\omega_{n}$, which is omitted for brevity of discussion. Substituting the expressions of ${\cal D}_{R0}$ and ${\cal D}_{i0}$ into (\ref{eq:DRB}), we have 
\begin{equation}
{\cal D}_{RB} =\frac{{\cal D}_{R0}}{1-{\cal D}_{R0}\left(\sum_{i}|c_{i}|^{2}{\cal D}_{i0}\right)} = \frac{2\omega_{d}}{(i\omega_{n})^{2}-\omega_{d}^{2}-2\omega_{d}\left(\sum_{i}\frac{2\omega_{i}|c_{i}|^{2}}{(i\omega_{n})^{2}-\omega_{i}^{2}}\right)}.\label{eq:DRB2}
\end{equation}
With the substitution $i\omega_{n}\rightarrow \omega+i\delta$ in the bath correlators, we have
\begin{equation}
\sum_{i}\frac{2\omega_{i}|c_{i}|^{2}}{\omega^{2}-\omega_{i}^{2}+i\delta}=\sum_{i}P\left(\frac{2\omega_{i}|c_{i}|^{2}}{\omega^{2}-\omega_{i}^{2}}\right)-i\pi\sum_{i}|c_{i}|^{2}\delta\left(\omega-\omega_{i}\right).\label{eq:bathterm1}
\end{equation}
The first term on the right hand side of the above equation is a summation of Cauchy principle values, and we denote this term as $\delta\omega_{d}$. This term generates a small shift of the cavity resonance to become $\widetilde{\omega}_{d}=(\omega_{d}^{2}+2\omega_{d}\delta\omega_{d})^{1/2}$ with $\delta\omega_{d}\ll\omega_{d}$. Using the definition of the noise spectrum $J(\omega)=\pi\sum_{i}|c_{i}^{2}|\delta(\omega-\omega_{i})$~\cite{bathreview} and the assumption $J(\omega)=\kappa$, we derive that 
\begin{equation}
\sum_{i}\frac{2\omega_{i}|c_{i}|^{2}}{\omega^{2}-\omega_{i}^{2}+i\delta}=\delta\omega_{d}-i\kappa.\label{eq:bathterm1}
\end{equation}
Finally, the detector correlator becomes
\begin{equation}
{\cal D}_{RB} = \frac{2\omega_{d}}{(i\omega_{n})^{2}-\widetilde{\omega}_{d}^{2}+i2\omega_{d}\kappa},\label{eq:DRB3}
\end{equation}
which corresponds to a cavity mode with frequency $\widetilde{\omega}_{d}\approx \omega_{d}$ and bandwidth $\sim\kappa$.


\begin{thebibliography}{99}
\bibitem{note1}The density operator of the electron gas can also be written as $\hat{O}_{S}(R_{0})=\sum_{R,\sigma}\hat{c}_{R\sigma}^{\dag} \hat{c}_{R\sigma} \delta_{R,R_{0}}$, where $\hat{c}_{R\sigma}$ ($\hat{c}_{R\sigma}^{\dag}$) is the annihilation (creation) operator of an electron with position $R$ and spin $\sigma$, $\delta_{R,R_{0}}=1$ only when $R=R_{0}$, and $\delta_{R,R_{0}}=0$ otherwise.
\end{thebibliography}
\end{document}